\documentclass[letterpaper]{aa}  

\usepackage{graphicx}

\usepackage{txfonts}

\begin{document}

\title{Downsizing of star-forming galaxies by gravitational processes}
\author{Hideaki Mouri\inst{1} \and Yoshiaki Taniguchi\inst{2}}

\offprints{H. Mouri}

\institute{Meteorological Research Institute, Nagamine, Tsukuba 305-0052, Japan\\
              \email{hmouri@mri-jma.go.jp}
              \and
           Physics Department, Graduate School of Science and Engineering, Ehime University, Matsuyama 790-8577, Japan}

\date{Received May 15, 2006; accepted August 5, 2006}

\authorrunning{Mouri \& Taniguchi}
\titlerunning{Downsizing of star-forming galaxies by gravitational processes}
 
\abstract
{There is observed a trend that a lower mass galaxy forms stars at a later epoch. This downsizing of star-forming galaxies has been attributed to hydrodynamical or radiative feedback processes that regulate star formation.}
{We explain the downsizing by gravitational processes alone, in the bottom-up scenario where galaxies evolve from subgalactic-scale objects.}
{Using the theory of Press and Schechter, as a function of galaxy mass, we study the peak epoch of formation of subgalactic-scale objects, i.e., gravitational collapse of subgalactic-scale density fluctuation.}
{The subgalactic-scale density fluctuation is found to collapse at a later epoch for a lower mass galaxy. The epoch is close to the peak epoch of star formation derived from observations.}
{The downsizing is inherent in gravitational processes of the bottom-up scenario. Within a region of the initial density field that is to evolve into a lower mass galaxy, subgalactic-scale fluctuation is of a smaller amplitude. Gravitational collapse of the subgalactic-scale density fluctuation and the subsequent onset of star formation occur at a later epoch for a lower mass galaxy. }


\maketitle

\section{Introduction}
\label{s1}

The current paradigm for the formation and evolution of galaxies is the bottom-up scenario based on the cold dark matter model. The cold dark matter retains significant small-scale fluctuation in the initial density field. Since the first objects are formed within dark matter halos that originate in gravitational collapse of the initial density fluctuation, these objects are not massive. They merge via gravitational clustering of the dark matter halos, evolve into low-mass galaxies, and then evolve into high-mass galaxies.

Observationally, there is a trend that is seemingly inconsistent with the bottom-up scenario, i.e., the ``downsizing'' of star-forming galaxies (Cowie et al. \cite{Cowie}; Heavens et al. \cite{Heavens}; Thomas et al. \cite{Thomas}; Treu et al. \cite{Treu}). Regardless of environment, a lower mass galaxy forms stars at a later epoch. In particular, a lower mass galaxy undergoes the peak of star formation at a later epoch. 

The existing explanations for the downsizing invoke hydrodynamical or radiative feedback processes that regulate star formation (De Lucia et al. \cite{Lucia}; see also the above references). For example, if the galaxy is not massive, supernova explosions could expel gas from the galaxy and thereby inhibit star formation until the fallback of the gas. However, here we demonstrate that the downsizing is in fact explainable by gravitational processes alone, using the theory of Press \& Schechter (\cite{Press}) for the bottom-up formation of dark matter halos.

\section{Definitions and equations}
\label{s2}

The standard values are assumed for the cosmological parameters: the Hubble constant $H_0 = 70$ km s$^{-1}$ Mpc$^{-1}$ ($h = 0.7$), matter density $\Omega_m = 0.3$, baryon density $\Omega_b = 0.04$, and cosmological constant $\Omega_{\Lambda} = 0.7$ (Spergel et al. \cite{Spergel}). We normalize the scale factor $a(t)$ to unity at the present epoch.

The mass density $\rho (\mbox{\boldmath $x$},t)$ and its average $\langle \rho \rangle = 3 \Omega _m H_0^2 / 8 \pi G$ lead to the density contrast $\delta (\mbox{\boldmath $x$},t) = \rho (\mbox{\boldmath $x$},t) / \langle \rho \rangle -1$. Here {\boldmath $x$} is the position in comoving coordinates and $\langle \cdot \rangle$ denotes an average. In the linear regime, the density contrast grows as
\begin{equation}
\label{e1}
\delta (\mbox{\boldmath $x$},t) = D(t) \delta (\mbox{\boldmath $x$}).
\end{equation}
The linear growth factor $D(t)$ is from Heath (\cite{Heath}):
\begin{equation}
\label{e2}
D(a) = \frac{5 \Omega_m}{2}
       \frac{H(a)}{H_0}
       \int ^a_0
       \frac{H_0^3da'}{a'^3 H(a')^3},
\end{equation}
with $H(a)/H_0 = [\Omega _m / a^3 + (1- \Omega _m - \Omega _{\Lambda}) / a^2 + \Omega _{\Lambda} ]^{1/2}$. The initial density contrast $\delta (\mbox{\boldmath $x$})$ is related to its Fourier transform $\tilde{\delta}(\mbox{\boldmath $k$})$ as
\begin{equation}
\label{e3}
\delta (\mbox{\boldmath $x$}) 
=
\frac{1}{(2 \pi)^{3/2}}
\int \tilde{\delta} (\mbox{\boldmath $k$}) 
\exp ( i \mbox{\boldmath $k$} \cdot \mbox{\boldmath $x$}) d \mbox{\boldmath $k$} .
\end{equation}
Since the initial density contrast is random Gaussian, its statistics are uniquely determined by the initial power spectrum $P(k)$ defined as
\begin{equation}
\label{e4}
\langle \tilde{\delta} (\mbox{\boldmath $k$} ) 
        \tilde{\delta} (\mbox{\boldmath $k$}')^{\ast} \rangle
=
(2 \pi )^3 \delta_{\rm D} (\mbox{\boldmath $k$} - \mbox{\boldmath $k$}') P(k) .
\end{equation}
Here $k = \vert \mbox{\boldmath $k$} \vert$ is the wavenumber, $\tilde{\delta} (\mbox{\boldmath $k$})^{\ast}$ is the complex conjugate of $\tilde{\delta} (\mbox{\boldmath $k$} )$, and $\delta_{\rm D}(\mbox{\boldmath $x$})$ is Dirac's delta function. 

The initial power spectrum originates in the Harrison-Zel'dovich spectrum $Ck$:
\begin{equation}
\label{e5}
P(k) = T(k)^2 Ck.
\end{equation}
We take the transfer function $T(k)$ from Bardeen et al. (\cite{Bardeen}):
\begin{eqnarray}
\label{e6}
T(k) &=& \frac{\ln (1+2.34q)}{2.34q} \\ 
     & & \times \ \ \left[ 1+3.89q+(16.1q)^2+(5.46q)^3+(6.71q)^4 \right] ^{-1/4} , \nonumber
\end{eqnarray}
with $q = k/ \Gamma h$ Mpc$^{-1}$ and $\Gamma = \Omega _m h \exp [- \Omega _b ( 1+\sqrt{2h}/ \Omega_m ) ]$ (Sugiyama \cite{Sugiyama}). Then, the normalization constant $C$ is set to be $7 \times 10^4$ on the basis of the root-mean-square linear density contrast smoothed over a radius $R_M = 8\,h^{-1}$\,Mpc observed for the present epoch, 0.8 (Spergel et al. \cite{Spergel}).

The smoothing is based on a top-hat window function: $W_M(\mbox{\boldmath $x$} ) = 3/4 \pi R_M^3$ for $| \mbox{\boldmath $x$}| \le R_M$ and $0$ for $| \mbox{\boldmath $x$}| > R_M$. Here $M$ is the mass for both baryonic and dark matter, and $R_M$ is the corresponding scale: $M = 4 \pi R_M^3 \langle \rho \rangle /3$. The Fourier transform of the smoothed density contrast $\delta_M (\mbox{\boldmath $x$})$ is
\begin{equation}
\label{e7}
\tilde{\delta}_M(\mbox{\boldmath $k$})
=
\tilde{W}_M(k) \tilde{\delta}(\mbox{\boldmath $k$}),
\end{equation}
with $\tilde{W}_M(k) = 3[ \sin (kR_M) - kR_M \cos (kR_M) ] / (kR_M)^3$. This differs from the usual Fourier transform of $W_M(\mbox{\boldmath $x$})$ by a factor $(2 \pi )^{3/2}$.

\section{Basic mechanism}
\label{s3}

Consider galaxies that have different masses at an epoch. This is consistent with the bottom-up scenario, although high-mass galaxies are few if the epoch is too early. The individual galaxies evolve from subgalactic-scale objects. Within such objects, star formation begins before the assembly of galaxy mass (see De Lucia et al. \cite{Lucia}). Using the theory of Press \& Schechter (\cite{Press}), we discuss that subgalactic-scale objects are formed later for a lower mass galaxy. Then, stars are formed later for a lower mass galaxy. This is just the downsizing of star-forming galaxies.\footnote{
Neistein et al. (\cite{Neistein}) gave a similar discussion based on a different approach. Their preprint was released just before the release of the second version of our preprint.}

For convenience, in our discussion, we equate formation of an object such as a galaxy with that of its dark matter halo. The same convention is used for the mass.

Press \& Schechter (\cite{Press}) considered that, even if objects have been formed at a scale, the larger scale density field remains linear. Regardless of substructure, the object formation occurs if the linear density contrast smoothed over that region reaches a certain critical value $\delta_{\rm c}$. Then, at an epoch $t$, objects with mass $M$ exist at all the positions where the smoothed linear density contrast $ D(t) \delta_M (\mbox{\boldmath $x$})$ is equal to the critical value $\delta_{\rm c}$. The positions with $D(t) \delta_M (\mbox{\boldmath $x$}) > \delta_{\rm c}$ belong to objects with higher masses. Here it is essential to specify the epoch $t$ because any object continuously increases its mass in the theory of Press \& Schechter (\cite{Press}).

The critical density contrast $\delta _{\rm c}$ is from gravitational collapse of a homogeneous sphere. In the Einstein-de Sitter universe with $\Omega _m = 1$ and $\Omega _{\Lambda} = 0$, the linear density contrast extrapolated to the epoch of complete collapse is $3 (12 \pi)^{2/3} / 20 \simeq 1.69$ (Gunn \& Gott \cite{Gunn}). This value is used as $\delta _{\rm c}$. The difference from the value for $\Omega _m = 0.3$ and $\Omega _{\Lambda} = 0.7$ is less than 1\% up to the present epoch (Percival \cite{Percival}).

\begin{figure}
\begin{center}
\resizebox{8.7cm}{!}{\includegraphics*[0.5cm,3cm][21.5cm,24.5cm]{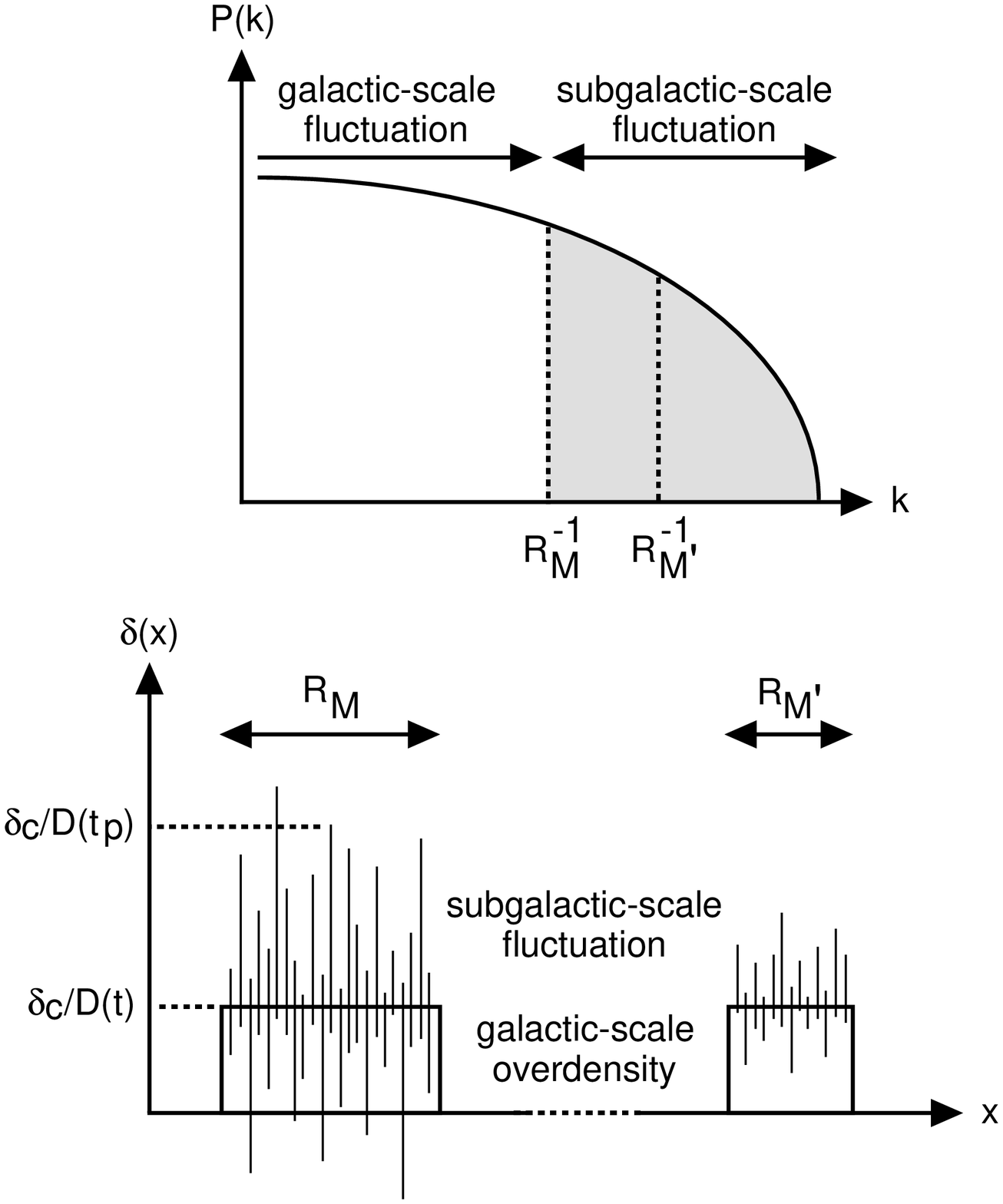}}

\caption[Fig1]{{\it Upper panel:} Initial power spectrum $P(k)$. The gray area indicates the power removed by the smoothing over the scale $R_M$. {\it Lower panel:} Initial density contrast $\delta(\mbox{\boldmath $x$})$. The scales $R_M$ and $R_{M'}$ correspond to galaxy masses $M$ and $M'$ ($R_M > R_{M'}$ for $M > M'$). \label{f1}}

\end{center}
\end{figure}

The basic mechanism of the downsizing is illustrated in Fig. \ref{f1}. We compare two regions of the initial density field that are to evolve into galaxies with different masses $M$ and $M'$ ($M > M'$) at an epoch $t$. The two regions respectively exist at any of the positions where $\delta_M(\mbox{\boldmath $x$})$ and $\delta_{M'}(\mbox{\boldmath $x$})$ are equal to $\delta_{\rm c}/ D(t)$. Superposed on these galactic-scale overdensities that have the same height, there are subgalactic-scale density fluctuations $\delta_{<M}(\mbox{\boldmath $x$}) = \delta(\mbox{\boldmath $x$}) - \delta_M(\mbox{\boldmath $x$})$ and $\delta_{<M'}(\mbox{\boldmath $x$}) = \delta(\mbox{\boldmath $x$}) - \delta_{M'}(\mbox{\boldmath $x$})$. Their variances $\langle \delta_{<M}^2 \rangle$ and $\langle \delta_{<M'}^2 \rangle$ are equal to the powers $ 4 \pi \int k^2 P(k) dk$ in the wavenumber ranges $k \gtrsim R_M^{-1}$ and $k \gtrsim R_{M'}^{-1}$. Since we set $R_M > R_{M'}$, we have $\langle \delta_{<M}^2 \rangle > \langle \delta_{<M'}^2 \rangle$. That is, the subgalactic-scale fluctuation is of a smaller amplitude for the lower mass galaxy. Hence, formation of subgalactic-scale objects, i.e., gravitational collapse of the subgalactic-scale fluctuation, occurs later for the lower mass galaxy. Since the gravitational collapse induces star formation, the star formation also occurs later for the lower mass galaxy. The lag of star formation behind the gravitational collapse is negligible, as far as the star formation occurs. This is because the dynamical timescale is small in dense dark matter halos that are formed at relatively high redshifts.

\section{Analytical calculation}
\label{s4}

For formation of subgalactic-scale objects with mass $M_{\ast}$, the peak epoch $t_{\rm p}$ is calculated as a function of galaxy mass $M$ at an epoch $t$. We set $M_{\ast} = 10^8$ or $10^9\,M_{\sun}$ as the minimum mass for significant star formation. The first stars are formed within minihalos with $10^6\,M_{\sun}$, but significant star formation occurs after the formation of halos with $\gtrsim 10^8\,M_{\sun}$ (Bromm \& Larson \cite{Bromm}; Kitayama \& Yoshida \cite{Kitayama}). We also set $M = 10^{10}$, $10^{11}$, $10^{12}$, or $10^{13}\,M_{\sun}$ as the typical galaxy mass.

The initial density field is smoothed over the scale $R_{M_{\ast}}$ that corresponds to the mass of subgalactic-scale objects $M_{\ast}$. From Eqs. (\ref{e4}) and (\ref{e7}), we have a new power spectrum
\begin{equation}
\label{e8}
P_{M_{\ast}}(k) = \tilde{W} _{M_{\ast}}(k)^2 P(k) .
\end{equation}
Eqs. (\ref{e3}), (\ref{e4}), (\ref{e7}), and (\ref{e8}) yield correlations among the galactic- and subgalactic-scale fluctuations $\delta_M(\mbox{\boldmath $x$})$ and $\delta_{<M}(\mbox{\boldmath $x$})$
\begin{eqnarray}
\label{e9}
\langle \delta_M^2 \rangle 
&=&4\pi \int ^{\infty}_0 k^2 \tilde{W}_M(k)^2 P_{M_{\ast}}(k) dk, 
\nonumber \\
\langle \delta_{<M}^2 \rangle
&=&4\pi \int ^{\infty}_0 k^2 [1-\tilde{W}_M(k)]^2 P_{M_{\ast}}(k) dk,  
\\
\langle \delta_M \delta_{<M} \rangle  
&=&4\pi \int ^{\infty}_0 k^2 \tilde{W}_M(k)[1-\tilde{W}_M(k)] P_{M_{\ast}}(k) dk. 
\nonumber
\end{eqnarray}
The correlation $\langle \delta_M \delta_{<M} \rangle$ has been ignored in Sect. \ref{s3}, but in fact it exists because $\tilde{W}_M(k)$ has extended oscillations. Since $\delta _M(\mbox{\boldmath $x$})$ and $\delta _{<M}(\mbox{\boldmath $x$})$ are random Gaussian, their joint probability is
\begin{eqnarray}
\label{e10}
p(\delta_M, \delta_{<M})
= \mbox{\hspace{6.5cm}}&&\\
\frac{1}{2\pi \sqrt{d}}
\exp \left( 
     \frac{ - \langle \delta_{<M}^2        \rangle \delta_M^2
            - \langle \delta_M^2           \rangle \delta_{<M}^2
            +2\langle \delta_M \delta_{<M} \rangle \delta_M \delta_{<M} }
          {2d}
     \right), & &\nonumber
\end{eqnarray}                    
with $d = \langle \delta_M^2 \rangle \langle \delta_{<M}^2 \rangle - \langle \delta_M \delta_{<M} \rangle ^2$. The probability for $\delta_M(\mbox{\boldmath $x$})$ is
\begin{equation}
\label{e11}
p(\delta_M) = \frac{1}{\sqrt{2\pi \langle \delta_M^2 \rangle}}
              \exp \left( - \frac{ \delta_M^2}{2\langle \delta_M^2 \rangle} \right) .
\end{equation}
We obtain the probability for the subgalactic-scale fluctuation $\delta_{<M}(\mbox{\boldmath $x$})$ within regions that are to evolve into galaxies with mass $M$ at an epoch $t$. This is the probability under the condition that $\delta_M(\mbox{\boldmath $x$})$ is equal to $\delta _{\rm c} /D(t)$:
\begin{equation}
\label{e12}
p_{M,t}(\delta_{<M}) = \frac{p[\delta_M = \delta_{\rm c}/D(t), \delta_{<M}]}
                      {p[\delta_M = \delta_{\rm c}/D(t)]}.
\end{equation}
The conditional probability $p_{M,t}(\delta_{<M})$ is Gaussian. Its average and variance are
\begin{eqnarray}
\label{e13}
\langle \delta_{<M} \rangle _{M,t}
&=& 
\frac{\langle \delta_M \delta_{<M} \rangle}{\langle \delta_M^2 \rangle}
\frac{\delta_{\rm c}}{D(t)}, \\
\label{e14} 
\langle ( \delta_{<M} - \langle \delta_{<M} \rangle )^2 \rangle _{M,t}
&=&
\sigma ^2 
=
\frac{d}{\langle \delta_M^2 \rangle}.
\end{eqnarray}
The average $\langle \delta_{<M} \rangle _{M,t}$ is smaller and hence less important than the $1\sigma$ fluctuation $\langle ( \delta_{<M} - \langle \delta_{<M} \rangle )^2 \rangle ^{1/2} _{M,t}$ for the parameter values in our calculation.


The epoch of gravitational collapse of the $1\sigma$ subgalactic-scale fluctuation is used as the peak epoch $t_{\rm p}$ of formation of subgalactic-scale objects within regions that are to evolve into galaxies with mass $M$ at an epoch $t$:
\begin{equation}
\label{e15}
\frac{\delta_{\rm c}}{D(t)}
+
\langle \delta_{<M} \rangle _{M,t}
+
\langle ( \delta_{<M} - \langle \delta_{<M} \rangle )^2 \rangle ^{1/2} _{M,t}
=
\frac{\delta_{\rm c}}{D(t_{\rm p})} .
\end{equation}
This equation represents the situation of Fig. \ref{f1}. The $1\sigma$ fluctuation corresponds to the typical height for local maxima of the subgalactic-scale fluctuation $\delta_{<M}(\mbox{\boldmath $x$})$ (Bardeen et al. \cite{Bardeen}, Fig. 2 for $\gamma = 0.5$).

The original theory of Press \& Schechter (\cite{Press}) has the ``cloud-in-cloud'' problem. They did not consider correlations among different mass scales. In particular, they did not consider whether an object belongs to another object with higher mass at the same epoch. Judging from the discussion of Nagashima (\cite{Nagashima}), since the correlation $\langle \delta_M \delta_{<M} \rangle$ is considered in Eq. (\ref{e15}), it is rather as accurate as the standard extension of the Press-Schechter theory (Bond et al. \cite{Bond}). There is an exception that objects with masses higher than $M$ are not considered in Eq. (\ref{e15}), but the consequences are not serious as seen below.

\begin{figure}
\begin{center}
\resizebox{8.7cm}{!}{\includegraphics*[1.2cm,13.5cm][19cm,26.3cm]{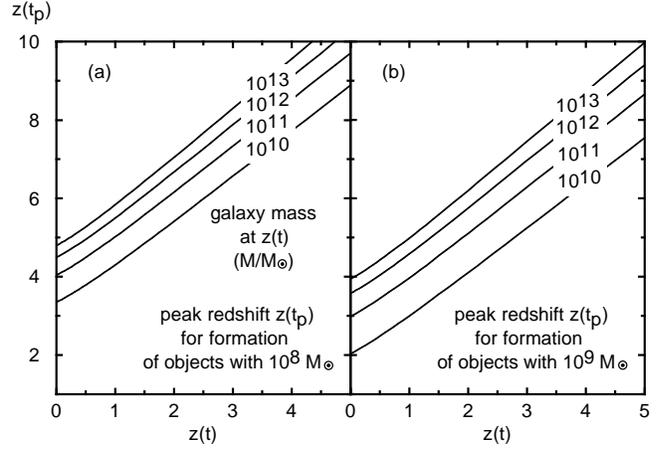}}
\caption[Fig2]{Peak redshift $z(t_{\rm p})$ of formation of subgalactic-scale objects calculated from Eq. (\ref{e15}). The galaxy mass $M$ at redshift $z(t)$ is $10^{10}$, $10^{11}$, $10^{12}$, or $10^{13}\,M_{\sun}$. The subgalactic-scale object mass $M_{\ast}$ is ($a$) $10^8\,M_{\sun}$ or ($b$) $10^9\,M_{\sun}$. \label{f2}}

\end{center}
\end{figure}

Our results are shown in Fig. \ref{f2}, where the epochs $t$ and $t_{\rm p}$ are converted into the redshifts $z(t)$ and $z(t_{\rm p})$. The peak redshift $z(t_{\rm p})$ of formation of subgalactic-scale objects is lower for lower galaxy mass $M$. We have thus reproduced the downsizing. The dependence of $z(t_{\rm p})$ on galaxy mass $M$ is more significant for lower galaxy mass $M$ and for higher subgalactic-scale object mass $M_{\ast}$. This is due to the narrower wavenumber range $R_M^{-1} \lesssim k \lesssim R_{M_{\ast}}^{-1}$ for the subgalactic-scale fluctuation.

The peak redshift $z(t_{\rm p})$ of formation of subgalactic-scale objects is consistent with the peak redshift of star formation derived from observations (Heavens et al. \cite{Heavens}; Thomas et al. \cite{Thomas}; Treu et al. \cite{Treu}), if the subgalactic-scale object mass $M_{\ast}$ is $10^8\,M_{\sun}$. The observations have uncertainties and discrepancies. Our best bet is that galaxies with $M \simeq 10^{12}$ to $10^{13}\,M_{\sun}$ at $z(t) = 0$ have the star formation peak at $z \simeq 4$ to 5.\footnote{
Supported by a semi-analytic model on a $N$-body simulation. The star formation peak is at $z \simeq 4.5$ and 5.0 for ellipticals with $M \simeq 10^{12}$ and $10^{13}\,M_{\sun}$ at $z(t) = 0$, respectively (De Lucia et al. \cite{Lucia}). Such ellipticals dominate the present population of galaxies with $M \ga 10^{12}\,M_{\sun}$. These $z$ values are almost equal to the corresponding $z(t_{\rm p})$ values in Fig. \ref{f2}$a$.} 
These $z$ values are close to the corresponding $z(t_{\rm p})$ values in Fig. \ref{f2}$a$. On the other hand, galaxies with $M \la 10^{11}\,M_{\sun}$ at $z(t) = 0$ have the star formation peak at $z \la 3$. The $z$ values are less than the corresponding $z(t_{\rm p})$ values. This is attributable to the cloud-in-cloud problem. The $z(t_{\rm p})$ values for $M \la 10^{11}\,M_{\sun}$ are overestimated because some objects with $M \la 10^{11}\,M_{\sun}$ are substructures of galaxies with $\ga 10^{12}\,M_{\sun}$. If we considered this effect, albeit necessarily in a complicated manner, the $z(t_{\rm p})$ values would become close to the peak redshifts of star formation.

The cloud-in-cloud problem also implies that some galaxies belong to clusters of galaxies. Since we ignore this implication, we ignore the dependence of the downsizing on environment. The dependence is actually negligible (Treu et al. \cite{Treu}; see also De Lucia et al. \cite{Lucia}). Although high-mass galaxies with old stellar populations tend to exist in clusters owing to the biased galaxy formation (Sect. \ref{s5}), high-mass field galaxies are equally dominated by old stellar populations.

\section{Discussion}
\label{s5}

The downsizing of star-forming galaxies, with a lower mass galaxy forming stars at a later epoch, is inherent in gravitational processes of the bottom-up scenario. Within a region of the initial density field that is to evolve into a lower mass galaxy, subgalactic-scale fluctuation is of a smaller amplitude (Fig. \ref{f1}). Gravitational collapse of the subgalactic-scale fluctuation and the subsequent onset of star formation occur later for a lower mass galaxy. As a function of galaxy mass, we have calculated the peak epoch of gravitational collapse of subgalactic-scale fluctuation (Fig. \ref{f2}). The peak epoch is consistent with the peak epoch of star formation derived from observations.

Our explanation for the downsizing differs from the existing explanation for the biasing, where gravitational collapse of small-scale density fluctuation occurs earlier if the large-scale overdensity in that region has a larger height (Kaiser \cite{Kaiser}; Bardeen et al. \cite{Bardeen}). We have demonstrated that, if the large-scale overdensity has a fixed height, its scale determines the epoch of gravitational collapse of small-scale density fluctuation (Fig. \ref{f1}). For the biased galaxy formation, the small and large scales correspond to galactic and supergalactic scales. For the downsizing of star-forming galaxies, the small and large scales correspond to subgalactic and galactic scales. The downsizing of star-forming galaxies is thereby independent of the biased galaxy formation.

Thus far, the downsizing has been explained by hydrodynamical or radiative feedback processes that regulate star formation. Although these processes are at work, they would not be essential. To the formation and evolution of galaxies, hydrodynamical and radiative processes are less essential than gravitational processes that solely explain the downsizing. Also, an observation suggests that the downsizing continuously extends from galaxies to clusters of galaxies where the star formation rate per unit mass at the present epoch is lowest (Feulner et al. \cite{Feulner}). Over a wide mass range from galaxies to clusters of galaxies, no processes are the same. The exception is gravitational processes, which apply to clusters of galaxies only if we consider subgalactic-scale density fluctuation superposed on cluster-scale overdensities in the initial field.

There is observed another trend that a lower mass galaxy has a longer history of star formation around its peak (Heavens et al. \cite{Heavens}; Thomas et al. \cite{Thomas}; see also De Lucia et al. \cite{Lucia}). This trend, obtained by plotting the star formation rate against epoch $t$, is explainable by the accelerated expansion of the universe and the resultant suppression of gravitational clustering of dark matter halos. The trend is not significant if the star formation rate is plotted against the linear growth factor $D(t)$.

The downsizing is also expected for clusters of galaxies undergoing galaxy formation. Within a region of the initial density field that is to evolve into a lower mass cluster, galactic-scale fluctuation is of a smaller amplitude. The galaxy formation, i.e., gravitational collapse of the galactic-scale fluctuation, occurs at a later epoch. Hence, a lower mass cluster is expected to contain younger galaxies. Further studies would be important to understand more about gravitational processes in the bottom-up scenario.

\begin{acknowledgements}
The authors are grateful to M. Enoki, N. Gouda, M. Morikawa, K. Okoshi, M. Shiraishi, and T. Yano for interesting discussions and to the referee for helpful comments.
\end{acknowledgements}

\end{document}